\documentclass[pdflatex,sn-mathphys-num]{sn-jnl}% Math and Physical Sciences Numbered Reference Style 
\usepackage{graphicx}%
\usepackage{multirow}%
\usepackage{amsmath,amssymb,amsfonts}%
\usepackage{amsthm}%
\usepackage{mathrsfs}%
\usepackage[title]{appendix}%
\usepackage{xcolor}%
\usepackage{textcomp}%
\usepackage{manyfoot}%
\usepackage{booktabs}%
\usepackage{algorithm}%
\usepackage{algorithmicx}%
\usepackage{algpseudocode}%
\usepackage{listings}%
\theoremstyle{thmstyleone}%
\theoremstyle{thmstyletwo}%

\theoremstyle{thmstylethree}%

\begin{document}

\title[Article Title]{Characterization of Tunnel Diode Oscillator for Qubit Readout Applications}


\author[1]{\fnm{Ivan} \sur{Grytsenko}}
\equalcont{These authors contributed equally to this work.}
\author[1,2]{\fnm{Sander} \sur{van Haagen}}
\equalcont{These authors contributed equally to this work.}

\author[1,3]{\fnm{Oleksiy} \sur{Rybalko}}

\author[1]{\fnm{Asher} \sur{Jennings}}

\author[1]{\fnm{Rajesh} \sur{Mohan}}

\author[1,4]{\fnm{Yiran} \sur{Tian}}

\author*[1,5]{\fnm{Erika} \sur{Kawakami}}\email{e2006k@gmail.com}

\affil[1]{\orgdiv{RIKEN Center for Quantum Computing}, \orgname{RIKEN}, \orgaddress{\street{2-1 Hirosawa}, \city{Wako}, \postcode{351-0198}, \state{Saitama}, \country{Japan}}}

\affil[2]{\orgdiv{Department of Applied Sciences}, \orgname{Delft University of Technology}, \orgaddress{\street{Lorentzweg 1}, \city{Delft}, \postcode{2628 CJ}, \country{The Netherlands}}}

\affil[3]{\orgdiv{Physics of Quantum Fluids and Crystals}, \orgname{B. Verkin Institute for Low Temperature Physics and Engineering of the National Academy of Sciences of Ukraine}, \orgaddress{\street{47 Nauky Ave.}, \city{Kharkiv}, \postcode{61103}, \state{Kharkiv}, \country{Ukraine}}}

\affil[4]{\orgdiv{Institute of Physics}, \orgname{Kazan Federal University}, \orgaddress{\street{16a Kremlyovskaya St.}, \city{Kazan}, \postcode{420008}, \state{Republic of Tatarstan}, \country{Russian Federation}}}

\affil[5]{\orgdiv{Cluster for Pioneering Research}, \orgname{RIKEN}, \orgaddress{\street{2-1 Hirosawa}, \city{Wako}, \postcode{351-0198}, \state{Saitama}, \country{Japan}}}


\abstract{We developed a tunnel diode oscillator and characterized its performance, demonstrating its potential applications in the quantum state readout of electrons in semiconductors and electrons on liquid helium. This cryogenic microwave source demonstrates significant scalability potential for large-scale qubit readout systems due to its compact design and low power consumption of only 1~$\mu$W, making it suitable for integration on the 10~mK stage of a dilution refrigerator. The tunnel diode oscillator exhibits superior amplitude stability compared to commercial microwave sources. The output frequency is centered around 140~MHz, commonly used for qubit readout of electrons in semiconductors, with a frequency tunability of 10~MHz achieved using a varactor diode. Furthermore, the phase noise was significantly improved by replacing the commercially available voltage source with a lead-acid battery, achieving a measured phase noise of -115~dBc/Hz at a 1~MHz offset.
}




\maketitle

\section{Introduction}\label{sec1}

One of the key features required to realize a fault-tolerant scalable quantum computer is the integration of energy-efficient and reliable qubit control and readout electronics. Recently, qubit control electronics have been successfully integrated using cryogenic Complementary Metal-Oxide-Semiconductor (CMOS) technology \cite{Van_Dijk2019-na,Bardin2019-gt,Pauka2021-st,Peng2022-xe,Pellerano2022-fz} , and superconducting Josephson-junctions~\cite{Howe2022-tf}. Here, we focus on developing the readout electronics using tunnel-diode oscillator~\cite{Chow1964-pb,Van_Degrift1975-sl} (TDO) circuits. TDO uses a tunnel diode as a negative resistance element~\cite{Reona1962-hv}, generating a microwave signal when connected to an LC circuit. Comparable to cryogenic CMOS devices (typical power consumption: 10~mW~\cite{Bardin2019-gt,Pauka2021-st,Peng2022-xe,Kang2022-ll,Lee2023-mx,Pellerano2022-fz}) and superconducting Josephson-junction circuits (100~$\mu$W~\cite{Howe2022-tf}), the TDO presented here consumes significantly less power, requiring only 1~$\mu$W.

\begin{figure}
    \centering
    \includegraphics[width=\linewidth]{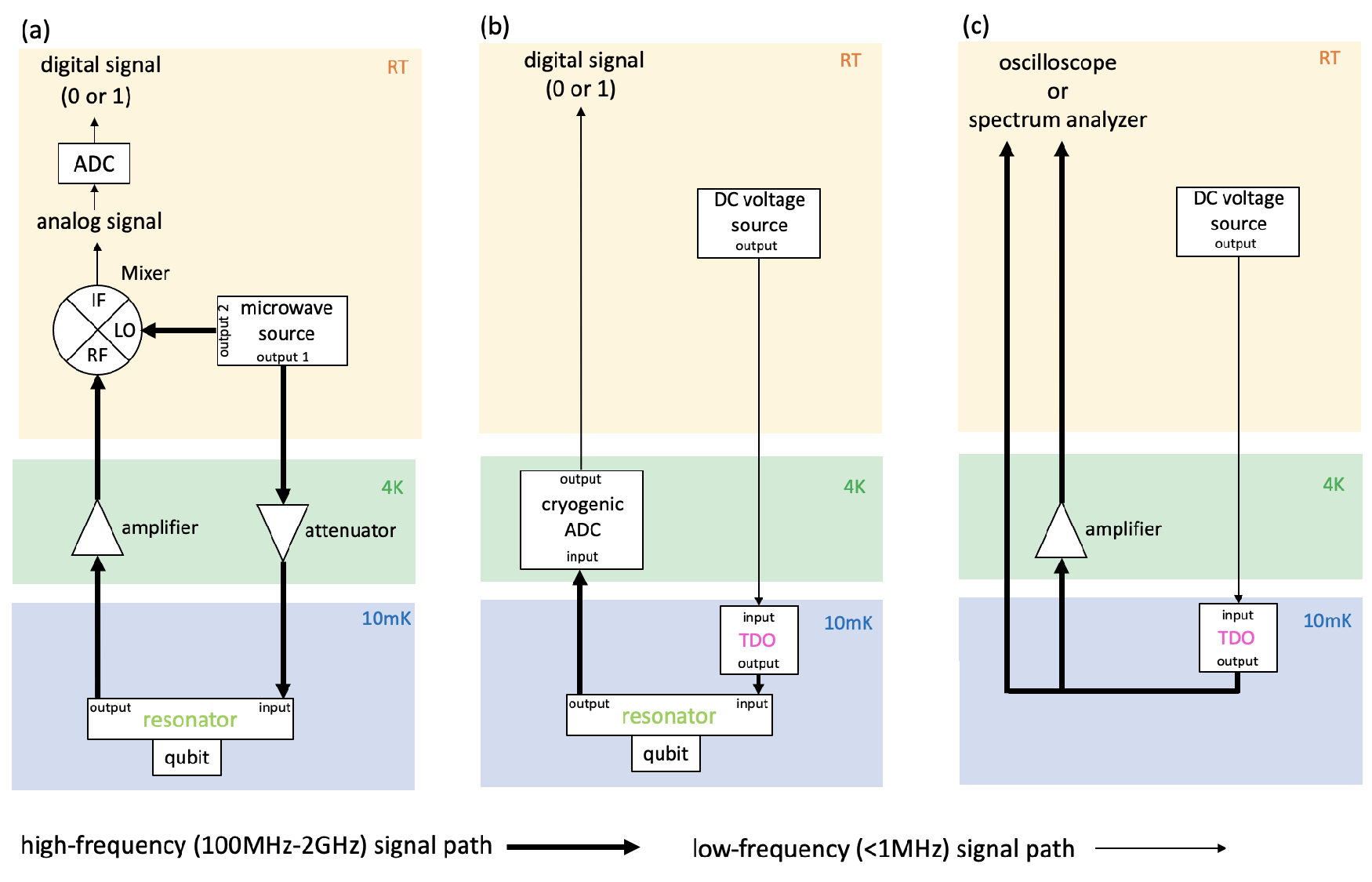}
    \caption{(a, b) Schematic diagrams illustrating dispersive qubit readout using a resonator: the conventional method in (a) and the tunnel-diode oscillator (TDO) method in (b). The bandwidth of the twisted pair of DC lines is limited to 1 MHz but is sufficient for transmitting the final calculation results to RT and powering the TDO from RT. (c) Schematic of the setup used to characterize the TDO. The characterization includes data measured both with and without the amplifier at 4~K.}
    \label{fig:motivation}
\end{figure}

This research aims to develop a technique that enables scalable qubit read-out schemes for quantum computers operating at cryogenic temperatures. The conventional method, illustrated in Fig.~\ref{fig:motivation}(a), is to generate a microwave signal with a room temperature (RT) microwave source and send it to a resonator dispersively coupled to the qubit at the 10~mK stage  (also called the mixing chamber, MC stage). The amplitude and phase of the reflected microwave signal depend on the qubit state, allowing us to determine the qubit state. This method is commonly used for qubits made from electrons in semiconductors~\cite{Hanson2007,Zwanenburg2013,Burkard2023-ai,Vigneau2023-yi,Urdampilleta2019-bo} and a similar method has been proposed for electrons on helium~\cite{Kawakami2023-vf}.

%The motivation of this research is to develop a technique that enables scalable qubit read-out schemes for quantum computers. Fig.~\ref{fig:motivation}(a) illustrates the conventional qubit readout method. This method is particularly relevant for qubits operating in cryogenic environments with temperatures around 10~mK, such as electrons used as qubits in semiconductors~\cite{Hanson2007,Zwanenburg2013,Burkard2023-ai,Vigneau2023-yi,Urdampilleta2019-bo}. A similar qubit readout method has also been theoretically proposed for electrons on helium~\cite{Kawakami2023-vf}.

In the conventional method, since the input signal is generated at RT, attenuators are placed at the 4~K stage to reduce thermal noise before it reaches the qubit. To prevent the reflected output signal from being overwhelmed by thermal noise at RT, it must be amplified at cryogenic temperatures. Cryogenic amplifiers typically have a lower noise figure, resulting in a higher signal-to-noise ratio. The resonant frequency of the resonator is typically in the range of 100~MHz to 2~GHz~\cite{Schoelkopf1998-nl,Colless2013-fc,Vigneau2023-yi}, and such high-frequency signals are transmitted through coaxial cables. A major scalability issue arises as the number of qubits increases because each qubit requires its own coaxial cable, which are relatively thick (feedthrough connector diameter: $\sim$1 cm, cable diameter: $\sim$1 mm) with attenuators and amplifiers also taking up the limited space inside a cryogenic refrigerator (diameter~$\sim$10-100~cm). Multiplexing can reduce the number of coaxial cables to some extent, but is limited by the resonator bandwidth~\cite{Hornibrook2014-pk}.

%In this method, a microwave signal is generated by a room-temperature (RT) microwave source and sent to a resonator dispersively coupled to the qubit. Attenuators are placed at 4K to reduce thermal noise from RT reaching the qubit. The amplitude and phase of the reflected microwave signal depend on the qubit state, allowing us to determine the qubit state through measurement. To prevent the signal from being overwhelmed by thermal noise at RT, it must be amplified before sending the signal to RT. The amplifier should be placed at cryogenic temperatures, as cryogenic amplifiers typically have a lower noise figure (NF), resulting in a higher signal-to-noise ratio. The resonant frequency of the resonator is typically in the range of 100~MHz to 2~GHz~\cite{Schoelkopf1998-nl,Colless2013-fc,Vigneau2023-yi}, and such high-frequency signals are transmitted through coaxial cables. A major scalability issue arises as the number of qubits increases because each qubit requires its own coaxial cable, which are relatively thick (diameter~$\sim$1~cm), while the space inside a cryogenic refrigerator is limited (diameter~$\sim$10-100~cm). Attenuators and amplifiers also take up space. Multiplexing can reduce the number of coaxial cables to some extent, but is limited by the resonator bandwidth~\cite{Hornibrook2014-pk}.

To address this challenge, we propose placing the microwave source closer to the qubits. By integrating the microwave source and qubits on the same circuit board, we can replace bulky coaxial cables with compact on-board transmission lines. This integration significantly reduces the circuit's size, making it more suitable for scaling up the number of qubits. Since the qubits must be operated at cryogenic temperatures (typically 10~mK), the microwave source must function at the same cryogenic stage. In this regard, the TDO we developed operates at the MC stage as the qubit. CMOS technology offers the advantage of high output power, operates at higher frequencies ($\sim$GHz), and performs complex functions; however, it must be placed at the 4 K stage due to its high power consumption.~\cite{Van_Dijk2019-na, Bardin2019-gt,Pauka2021-st,Peng2022-xe,Pellerano2022-fz}. It should be noted that, while not as an oscillator, CMOS has been used at the 10~mK stage to implement other circuits for quantum device control, such as multiplexers and DC bias circuits. The TDO is powered by a DC voltage supplied through DC lines. For these scenarios, an analog-to-digital converter (ADC) at cryogenic temperatures~\cite{Braga2024-bx} (Fig.~\ref{fig:motivation}(b)) can be placed to remove the need for coaxial cables on the resonator output lines as well. This approach would enable faster feedback for quantum error correction~\cite{Fowler2012} by eliminating the need to transmit signals to RT, requiring only the final computation results to be sent back. Additionally, placing the ADC at cryogenic temperatures removes the need for amplifiers, as thermal noise from RT would no longer be a concern.

In this work, we focus on the characterization of the TDO without utilizing an ADC or a resonator coupled to qubits (Fig.~\ref{fig:motivation}(c)). This initial step allows us to evaluate the performance and stability of the TDO as a standalone device under cryogenic conditions, providing a foundation for future integration with qubits. For qubit applications, microwaves are used for qubit state readout and manipulation. Amplitude stability is particularly important for readout~\cite{Vigneau2023-yi}, while phase noise also plays a crucial role in manipulation~\cite{Ball2016-zn}. Therefore, in this work, we characterized both amplitude stability and phase noise to evaluate the performance of the TDO in this context.

\section{Tunnel Diode Oscillator}\label{sec2}

We use the commercial BD-6 Ge-backward tunnel diode (American Microsemiconductor, Inc.) because it has the lowest power consumption among the commercially available options. TDOs with backward tunnel diodes demonstrate frequency stability of 0.001 ppm~\cite{Van_Degrift1975-sl,Van_Degrift1981-pe}. Leveraging this precision, they have been employed in the past to investigate various physical properties, such as paramagnetic susceptibility in salts~\cite{Clover1970-uo}, penetration depth in superconductors~\cite{Hashimoto2010-ao,Fletcher2009-qw}, and the melting point of $^3$He~\cite{Mikheev1989-wf}. These results were obtained with long integration times using a frequency counter, whereas qubit readout demands precise measurements within shorter durations. This emphasizes the need to understand the frequency and amplitude stability achievable over shorter integration times. Furthermore, the oscillation frequencies, typically around 10 MHz~\cite{Clover1970-uo,Al-Harthi2007-nr,Hashimoto2010-ao,Fletcher2009-qw,Srikanth1999-vo,El-Basit2020-pr}, are relatively low for qubit readout applications. In the early stages of testing qubit readout, it is also desirable to have the capability to externally tune the frequency. To address this issue, we developed a TDO operating at around 140 MHz, utilizing a microfabricated superconducting spiral inductor. This frequency has reached the lower bound among experiments~\cite{Verduijn2014-px} where the electron transition is measured through dispersive coupling to a resonator, making it applicable for qubit readout. By incorporating a varactor diode as a variable capacitor, we achieved a frequency tunability of 10 MHz without altering the amplitude. Additionally, this oscillator might also be used as a low-power, low-phase-noise clock source for cryo-ADCs.

\begin{figure}[H]
    \centering
    \includegraphics[width=\linewidth]{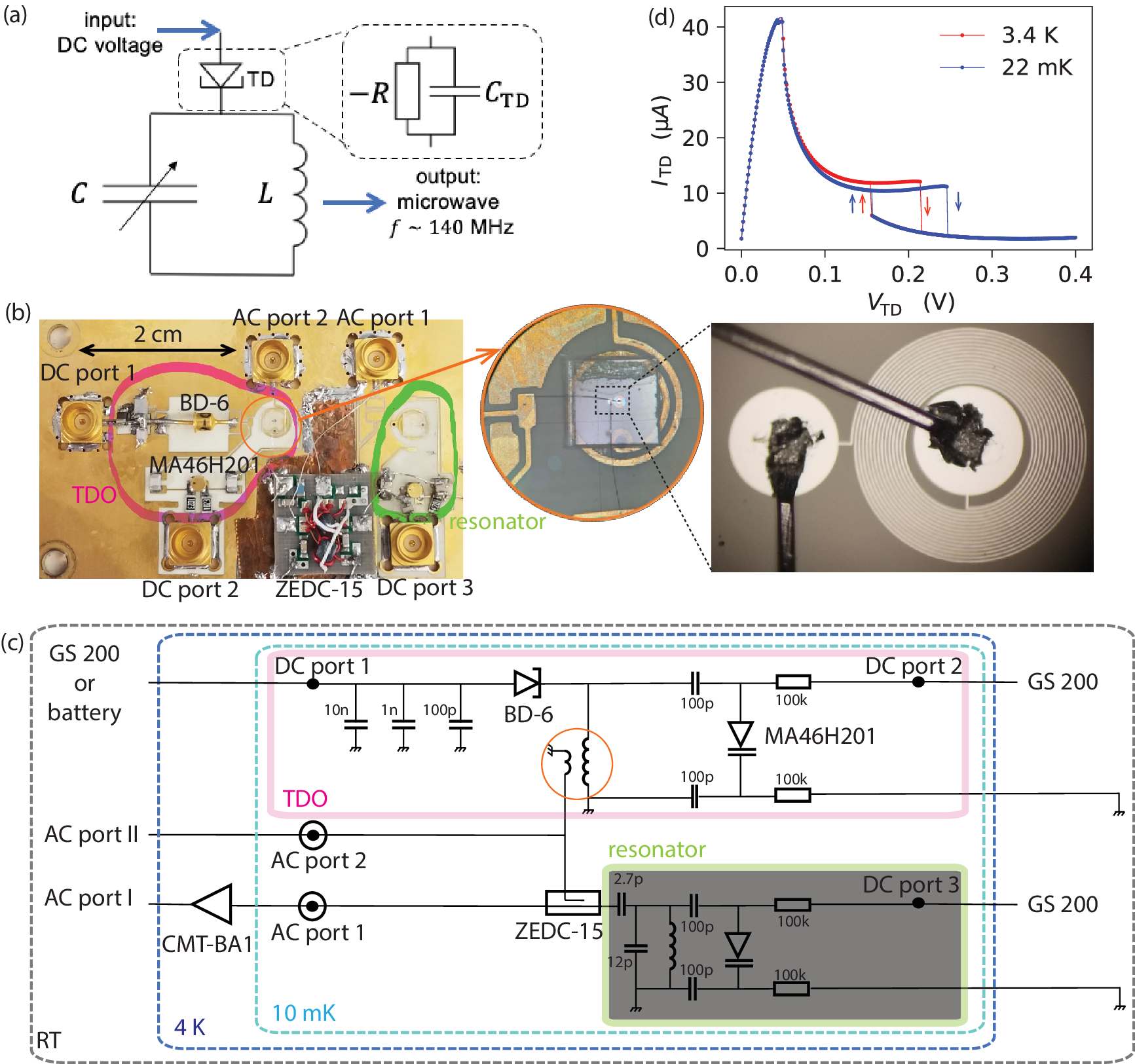}
    \caption{(a) Conceptual circuit illustration of the TDO. The tunnel diode (TD) can be modeled as a negative resistance \( -R \) and a capacitance \( C_\mathrm{TD} \).  The input voltage applied to the TD is converted into and output as a microwave signal with a frequency of  $\sim$140 MHz. (b) A photograph of the developed circuit. The pink box encloses the TDO, and the green box the resonator. The orange circle highlights a microfabricated 15-turn spiral inductor on a sapphire substrate, positioned on top of a 1.5-turn pickup coil, enabling a 10\% signal to be extracted from the TDO in terms of voltage. The pickup coil and the TDO are connected via a coupler (ZEDC-15). In this study, the resonator is detuned from its resonance, behaving as an open-ended element in the circuit. An enlarged photograph of the microfabricated 15-turn spiral inductor on a sapphire substrate is also shown. The diameter of its boding pad is 176 $\mu$m. (c) Circuit diagram of the device shown in (a). See the main text for details of the circuit components. The signal output from the TDO is measured at RT using a digital oscilloscope or a spectrum analyzer via AC port I or AC port II. For comparison, a signal from a commercial microwave source is sent to the cryogenic device through AC port II, and the reflected signal from the resonator is measured at RT using a digital oscilloscope or a spectrum analyzer via AC port I. The resonator is shaded in gray because its resonant frequency is set far outside the TDO's operating range, allowing it to function as a high-impedance element and remain inactive. (d) The I-V curve of the tunnel diode BD-6 at the mixing chamber stage temperature (MCT) of 22~mK and 3.4~K. Hysteresis is observed only during oscillation, depending on the sweep direction of the DC bias $V_\mathrm{TD}$ applied to the tunnel diode (indicated by arrows). When the bias is increased in the forward direction, oscillation occurs in the range of 0.04 V to 0.245 V, whereas in the reverse direction, the oscillation range is from 0.15 V to 0.04 V. }
    \label{fig:circuit}
\end{figure}

Fig.~\ref{fig:circuit}(a) shows a conceptual circuit illustration explaining how a tunnel diode (TD) implements oscillations. When a TD is connected in an LC resonant circuit, the negative resistance (represented by $-R$ in Fig.~\ref{fig:circuit}(a)) cancels out the resistive losses, allowing sustained oscillations. The oscillation frequency is determined by the LC resonance condition:  
\begin{equation}
    f = \frac{1}{2\pi \sqrt{L(C + C_{\mathrm{TD}})}} \label{eq:freq}
\end{equation}
where \( C_{\mathrm{TD}} \) represents the capacitance of the TD. \( C \) is the sum of the parasitic capacitance of the circuit board and the capacitance of the varactor diode, and its capacitance value is tunable by changing the voltage applied to the varactor diode, \( V_{\mathrm{VD}} \). To achieve high-frequency oscillations, minimizing parasitic capacitance in both the circuit and its components is crucial. Additionally, a high-Q inductor is important, as the self-resonant frequency should be higher than the oscillation frequency.

It is important to carefully select suitable BD-6 tunnel diodes for two reasons: first, to ensure that they exhibit negative resistance at low temperatures, and second, to minimize parasitic capacitance. Although  the I-V characteristics measured at RT show little variation among different BD-6 tunnel diodes, some lose their negative resistance at low temperatures (see Appendix~\ref{secA:Indiv_diff}). Therefore, screening using low-temperature I-V measurements is necessary. BD-6 tunnel diodes also exhibit significant variation in parasitic capacitance. We measured 62 different BD-6 tunnel diodes and found that their parasitic capacitance at RT ranged from \( C_\mathrm{TD} = 1.3 \) pF to 29.4 pF. The tunnel diode used in this study has a parasitic capacitance of approximately $C_\mathrm{TD}=$7 pF (Table.~\ref{tab:capacitance}).

For high-frequency oscillation, the inductance should simultaneously achieve a high Q-factor and a resistance lower than the 5 k$\Omega$ negative resistance of the tunnel diode. Therefore, we  microfabricated 15-turn Nb spiral coil as an inductor, which was separately measured to be 95~nH at 4~K. Relatively high frequencies, around 100~MHz, have been achieved using a different approach based on a toroidal LC resonator~\cite{Van_Degrift1974-ci}. However, due to its bulky nature, this approach is unsuitable for scalable qubit applications.

Fig.~\ref{fig:circuit}(b,c) shows the details of the developed circuit. This circuit board is thermally anchored to the 10~mK stage, which is referred to as the Mixing Chamber (MC) stage. The part enclosed by the pink line represents the TDO, which consists of the Nb spiral inductor and a variable capacitance formed by a varactor diode (MA46H201, MACOM). The varactor diode allows tuning of the LC resonator's capacitance, thereby varying the oscillation frequency of the TDO. The varactor diode is biased via DC port 2 using a DC voltage source (Yokogawa GS200). The oscillation signal is coupled to the pickup coil with a coupling ratio of 15:1.5.  In previous work~\cite{Clover1970-uo,Al-Harthi2007-nr,Hashimoto2010-ao,Fletcher2009-qw,Srikanth1999-vo,El-Basit2020-pr}, the signal was extracted capacitively from the same line used to provide the DC bias to the tunnel diode. Instead, in this work, the signal is extracted inductively. This approach ensures that the DC bias line is used exclusively for providing the DC bias, thereby preventing the TDO oscillation from being affected by filters and other components further along the line. As a result, a more stable DC bias can be supplied to the tunnel diode. The tunnel diode is biased via DC port 1 using either the Yokogawa GS200 or a lead-acid battery. After the pickup coil, the signal is sent to AC port II and measured at RT. It is also routed through a directional coupler (ZEDC-15, Mini-Circuits). The coupling from the coupling port to the input/output port of the ZEDC-15 is 20~dB, as measured separately at the 4~K stage. The signal from the coupler is amplified at the 4~K stage using a cryogenic amplifier (CMT-BA1, Cosmic Microwave Technology) and measured at RT with a spectrum analyzer or a digital oscilloscope.

The part enclosed by the green line in Fig.~\ref{fig:circuit}(b, c) represents the resonator, which can be coupled to a qubit. In this work, the resonator's resonant frequency is set far outside the TDO's operating range, allowing it to act as a high-impedance element and remain unused. 

The operating point of the tunnel diode BD-6, where it exhibits negative resistance and oscillates, is approximately \( V \approx 0.1 \, \text{V} \) and \( I \approx 10 \, \mu\text{A} \) (Fig.~\ref{fig:circuit}(d)), resulting in a power consumption of about \( 1 \, \mu\text{W} \). When oscillations occur, the DC bias conditions are slightly modified due to the self-rectification of alternating currents by the TD. This explains the hysteresis observed in the I-V curve in Fig.~\ref{fig:circuit}(d) depending on the sweep direction of $V_\mathrm{TD}$ ~\cite{Van_Degrift1981-pe}.

\section{Frequency and power tunability}\label{sec:freq_power_tunability}

\begin{figure}[H]
    \centering
    \includegraphics[width=0.8\linewidth]{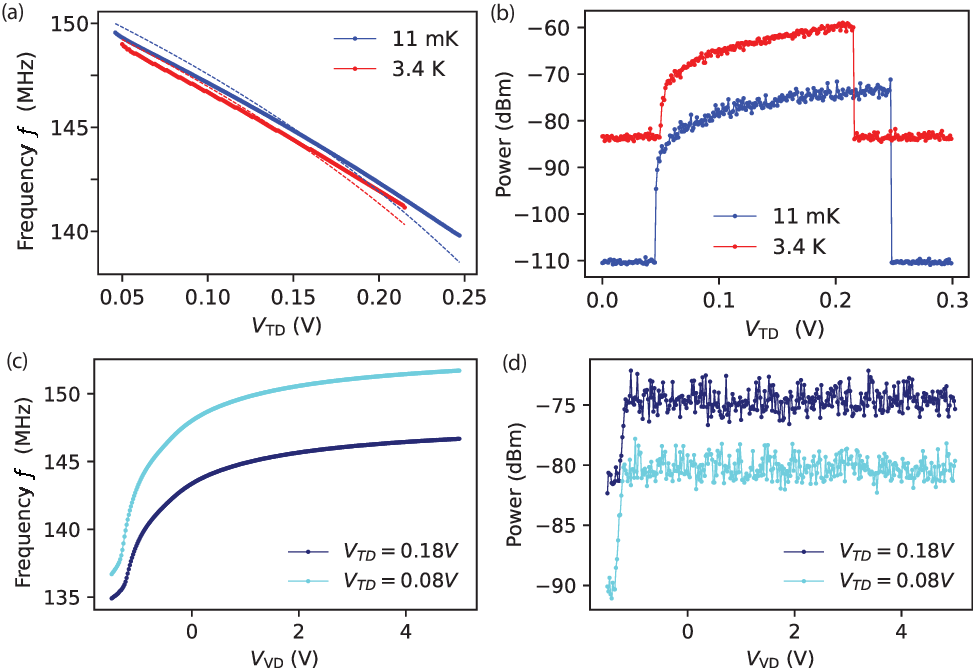}
    \caption{The signal from the TDO obtained through AC port II was measured using a spectrum analyzer. (a,b) $V_\mathrm{VD}$ is set to 0. The sweep direction of $V_\mathrm{TD}$ is forward. 
(a) Output frequency as a function of $V_\mathrm{TD}$ at MCT = 11~mK and 3.4~K. The dashed lines represent the fitted curves, which account for changes in depletion layer thickness induced by varying $V_\mathrm{TD}$. See the main text for details on the fitting procedure. 
(b) Output power as a function of $V_\mathrm{TD}$ at MCT = 11~mK and 3.4~K. The difference in background levels is attributed to the use of different spectrum analyzers for measurements at MCT = 11~mK and 3.4~K. 
(c) Output frequency as a function of $V_\mathrm{VD}$ at MCT = 11~mK.  
(d) Output power as a function of $V_\mathrm{VD}$ at MCT = 11~mK. It remains constant for $V_\mathrm{VD} > -1.3$~V.}
    \label{fig:freq_tunability}
\end{figure}

In this section, we demonstrate that the output power and frequency can be controlled by varying the voltage applied to the TD and that the frequency can be independently adjusted by varying the voltage applied to the varactor diode. The frequency variation due to $V_\mathrm{TD}$ is shown in Fig.~\ref{fig:freq_tunability}(a). The capacitance of the TD, \(C_{\mathrm{TD}}\), is voltage-dependent because the distance of the depletion layer in the p-n junction changes as a function of $V_\mathrm{TD}$. It is given by~\cite{Chow1964-pb} 
\begin{equation}
    C_{\mathrm{TD}} = C_0 \left( 1- V_{\mathrm{TD}}/V_{\mathrm{d}} \right)^{-1/2}, \label{eq:CTD}
\end{equation}
where $V_d=0.5$~V is the diffusion potential (estimated from a measurement) and $C_0$ is the capacitance of the TD when $V_\mathrm{TD}=0$. Eq.~\ref{eq:freq} with Eq.~\ref{eq:CTD} has been fitted, as shown by the dashed lines in Fig.~\ref{fig:freq_tunability}(a). The capacitance slightly decreases when the MCT is lowered from 3.4 K to 11 mK. The fitting results yield $C = 5.8$~pF and $C_0 = 5.7$~pF at MCT = 11~mK, and $C = 5.9$~pF and $C_0 = 5.8$~pF at MCT = 3.4~K. Fig.~\ref{fig:freq_tunability}(b) demonstrates that the output power can be tuned by approximately 10 dB by adjusting \( V_\mathrm{TD} \), as the operating point of the TD shifts along its I-V curve.  The microwave power that would be delivered to the resonator is around -90~dBm, which is comparable to the optimized power range for qubit readout of electrons used as qubits in semiconductors~\cite{Gonzalez-Zalba2015}.

Fig.~\ref{fig:freq_tunability}(c) shows the frequency tunability as a function of $V_\mathrm{VD}$. Considering the additional changes due to \( V_{\mathrm{TD}} \), we observe a total frequency tunability of 10~MHz. From this, we estimate that \( C \) varies from 7.9~pF to 5.3~pF as \( V_{\mathrm{VD}} \) is changed from -1.5~V to 5~V (Table. \ref{tab:capacitance}).

Fig.~\ref{fig:freq_tunability}(d) shows that for $V_\mathrm{VD} > -1.3$~V, the output power remains unaffected by $V_\mathrm{VD}$, indicating that the resistive component of the varactor remains unchanged (leakage current stays significantly small within this voltage range as seen in Fig.~\ref{fig:varactor}). Fig.~\ref{fig:freq_tunability}(d) shows that for \( V_\mathrm{VD} > -1.3 \)~V, the output power remains unaffected by \( V_\mathrm{VD} \), indicating that the resistive component of the varactor remains unchanged. This is consistent with the fact that the leakage current was measured to be significantly low within this voltage range, as shown in Fig.~\ref{fig:varactor}.
Please note that we have observed a change in output power when $V_\mathrm{VD}$ is varied within the same voltage range at RT when using a similar circuit with a normal coil inductor. We attribute this difference to the fact that the leakage current becomes significantly suppressed at low temperatures. 

\begin{table}[h]
       \caption{The table presents capacitance values under different voltage conditions at MCT = 11~mK. At 3.4~K, the capacitance values are higher by approximately 0.1~pF compared to those at 11~mK in all cases.}
       \centering
    \renewcommand{\arraystretch}{1.3}  % Adjust row height for better readability
    \begin{tabular}{c|ccc|ccc}
        \hline
        & \multicolumn{3}{c|}{\( C_{\text{TD}} \)} & \multicolumn{3}{c}{\( C \)} \\
        \hline
        Voltage  (V) & 
        \( V_{\text{TD}}=0 \) & 
        \( V_{\text{TD}}=0.08 \) & 
        \( V_{\text{TD}}=0.18 \) & 
        \( V_{\text{VD}}=-1.5 \) & 
        \( V_{\text{VD}}=0 \) & 
        \( V_{\text{VD}}=5 \) \\
        \hline
        Capacitance  (pF)& 
        5.7 & 6.3 & 7.2 & 7.9 & 5.8 & 5.3 \\
        \hline
    \end{tabular}
    \label{tab:capacitance}
\end{table}

\section{Phase noise}\label{sec:phase_noise}

\begin{figure}[H]
\centering
\includegraphics[width=0.9\linewidth]{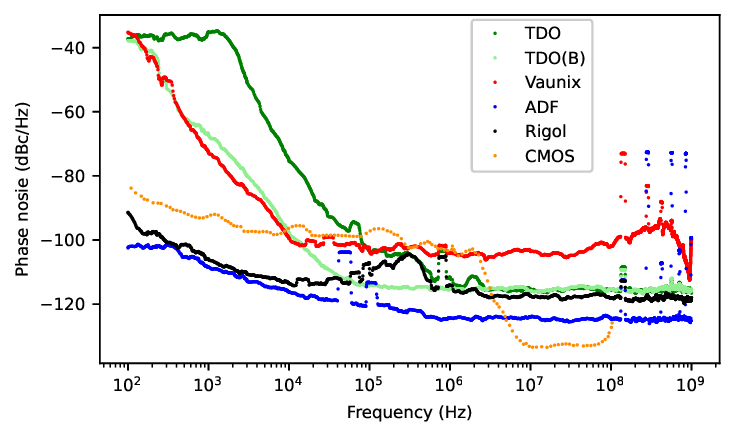}
\caption{Measured phase noise relative to the carrier frequency (around 141.8 MHz) as a function of offset frequency with a bandwidth of 1~Hz. The phase noise of four different commercial microwave sources is shown: Rigol DSA815 (black), Analog Devices ADF4351 (blue), Vaunix Lab-Brick LMS-451D (red), along with the TDO powered by Yokogawa GS200 DC source (green) and by a lead-acid battery (light green). For comparison, the phase noise generated by a CMOS device at a carrier frequency of 600 MHz is also shown (orange)~\cite{Xue2023-cj}. The jumps observed below \(2 \times 10^5\) Hz are likely caused by noise from the pulse tube of the dilution refrigerator. The jumps around \(10^6\) Hz are attributed to interference from a nearby radio station transmitting at 810 kHz (Appendix~\ref{secA:DC_source}). The jumps above \(10^8\) Hz are due to harmonics of the carrier frequency.}
\label{fig:phase_noise}
\end{figure}

In this section, we report the phase noise of the signal from the TDO operating at MCT=11~mK through AC port I using a spectrum analyzer (R\&S FSV3030). Additionally, we sent the signal from the commercial microwave source through AC port II and measured the phase noise of the reflected signal through AC port I. The results are compared and shown in Fig.~\ref{fig:phase_noise}. Note that the phase noise of the commercial microwave source, measured directly on the spectrum analyzer without routing it to the device at low temperature, showed almost identical results. This indicates that the measured phase noise originates from the microwave sources and not from the device at low temperature.

We found that using a lead-acid battery to bias the TDO instead of the Yokogawa GS200 improves the phase noise as shown in Fig.~\ref{fig:phase_noise} (see also Sec.~\ref{secA:DC_source} for details).  The measured phase noise of the TDO, powered by a battery, was $-115$~dBc/Hz at a 1~MHz offset, demonstrating comparable or superior performance to CMOS devices operating at the 4~K stage in a similar carrier frequency range~\cite{Xue2023-cj,Yang2019-bb,Lee2023-mx}. While CMOS devices achieve better phase noise in the lower- and higher-frequency ranges, the TDO provides better performance within the mid-frequency range. Moreover, with the battery as its power source, the TDO's phase noise is either comparable to or outperforms that of a commercial microwave source (Vaunix Lab-Brick LMS-451D) across the entire measured frequency range. 

The TDO has comparable or higher phase noise in the low-frequency range compared to commercial microwave sources. To improve this, enhancing the stability of the DC bias could help mitigate noise fluctuations. Additionally, the phase noise might be caused by two-level systems in the device, suggesting that improvements in the tunnel diode itself may be necessary. Another potential improvement is integrating a feedback mechanism to actively stabilize the frequency or implementing advanced filtering techniques to suppress unwanted noise components.

\section{Amplitude stability}\label{sec5}

\begin{figure}[H]
\centering
\includegraphics[width=0.9\linewidth]{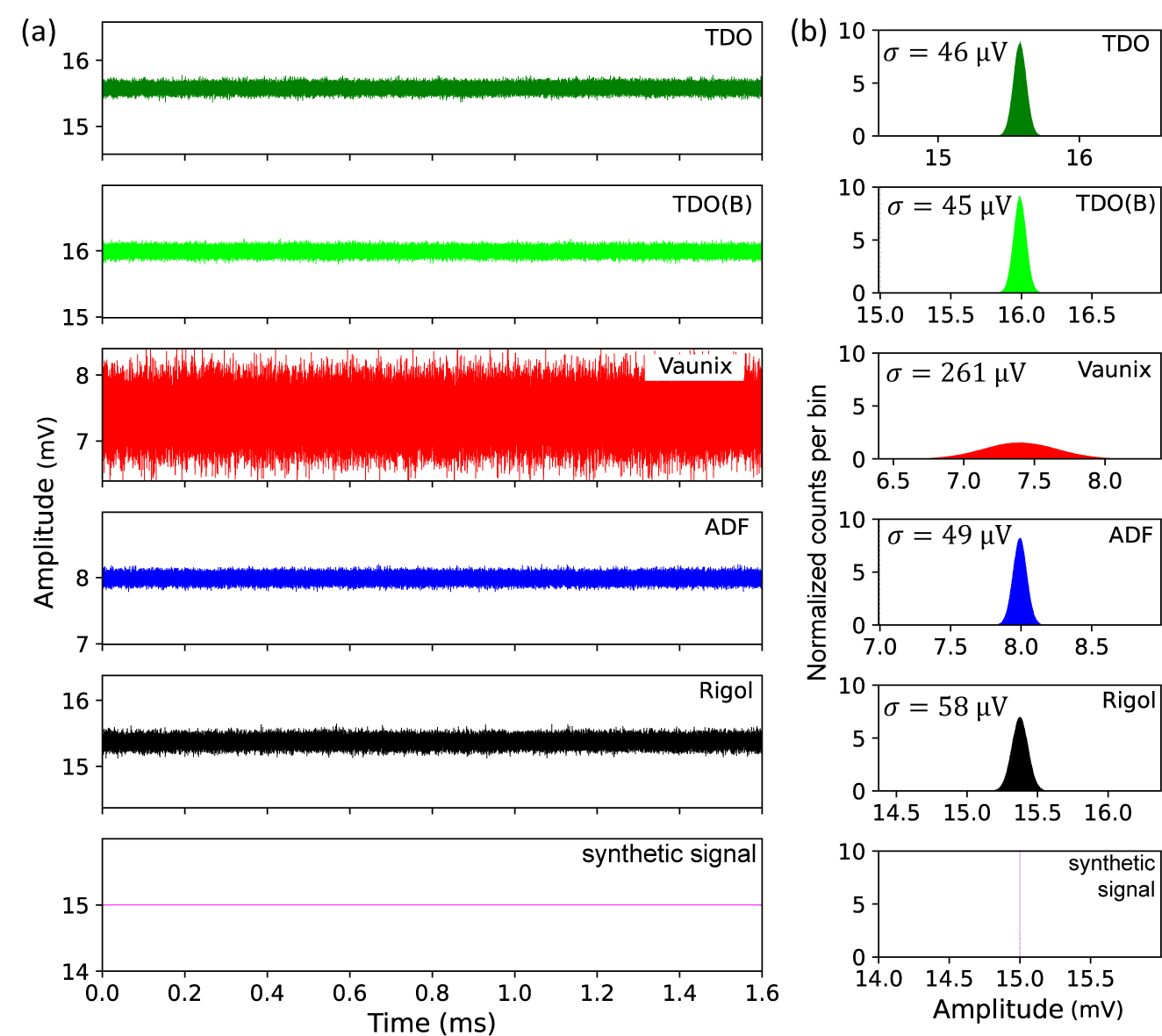}
\caption{The amplitude range is set to 2~mV for all the data for comparison. (a) The amplitude fluctuations over 1.6~ms are shown; see the main text for details. (b) Histograms of the amplitudes with 5000 bins and their standard deviations are also shown.}
\label{fig:amplitude-stability}
\end{figure}

In this section, we measured the amplitude of the signal from the TDO operating at MCT=11~mK through AC port I. The TDO generated a signal at 141.8~MHz with $V_\mathrm{TD}=0.18$~V and $V_\mathrm{VD}=-0.5$~V. The signal, recorded at a sampling rate of 20~GS/s for 1.6~ms, was passed through a digital bandpass filter centered at the carrier frequency with a 50~MHz bandwidth (see Appendix~\ref{secA:time-domain}). The amplitude was then extracted using a Hilbert transform and is plotted in Fig.~\ref{fig:amplitude-stability}(a), and its histogram is shown in Fig.~\ref{fig:amplitude-stability}(b). Similar to the phase noise measurement, a signal from a commercial microwave source was sent through AC port II, and the amplitude stability of the reflected signal, measured via AC port I, is shown in Fig.~\ref{fig:amplitude-stability} for comparison. For validation of our analytical method, a synthetic sine signal at 141.8~MHz generated is also plotted. Regardless of whether a lead-acid battery or the Yokogawa GS200 was used as the DC source, the amplitude stability of the TDO was superior to that of any commercial microwave source used in this study.

As with the phase noise measurements, the amplitude stability of the commercial microwave source, measured directly without routing it to the device at low temperature, showed identical results to those obtained when the signal was passed through the cryogenic device. This indicates that the observed amplitude stability is determined by the microwave sources themselves and not influenced by the device at low temperatures.

\section{Discussion}\label{sec:discuss}

We do not yet know the origin of the amplitude fluctuations, and identifying it will be the subject of future work. Since the oscilloscope used in this study has an 8-bit resolution, its resolution of \(1/2^8 = 0.39\%\) is comparable to the observed fluctuations of \(45 \, \mu\text{V}/15 \, \text{mV} = 0.3\%\). This suggests that the fluctuations might arise from the resolution limit of the oscilloscope rather than being intrinsic to the TDO. To investigate this further, improvements in the measurement setup, such as using higher-resolution ADCs, will be necessary. For qubit readout, particularly for electrons in semiconductors, where the qubit state is determined by observing the amplitude distribution~\cite{Burkard2023-ai,Vigneau2023-yi,Urdampilleta2019-bo}, it is unlikely that the TDO we developed would result in worse readout fidelity compared to the commercial microwave sources used here. However, looking forward, it will be necessary to evaluate the level of amplitude stability required for qubit readout using actual qubits.

Although the current study focuses on qubit readout, the TDO could potentially be utilized for qubit operation. However, its phase noise is currently not superior to that of commercial sources and could become a limiting factor if used for qubit operation~\cite{Ball2016-zn}. Further studies are needed to determine the phase noise requirements for qubit operation. Moreover, for qubit operation, the frequency of 140~MHz is too low. For operations involving electron spin states, GHz-range frequencies would be required. To achieve oscillation at higher frequencies, it is essential to reduce the parasitic capacitance of the TD and the circuit board. The parasitic capacitance of tunnel diodes exhibits significant variation, which is unlikely to be solely due to the pn-junction capacitance. Instead, this variation may originate from the diode packaging. Therefore, rather than relying on commercially available tunnel diodes, designing a tunnel diode specifically optimized for qubit readout or operation applications would be more effective. Minimizing the parasitic capacitance on the circuit board is also important. One possible approach is to fabricate the entire circuit board using microfabrication techniques. Another approach is to replace bulky varactor diodes with variable capacitors made from ferroelectric materials, such as strontium titanate (STO)~\cite{Apostolidis2024-kr}. Ferroelectric materials like STO not only allow for variable capacitance but also eliminate leakage currents, reducing circuit losses and providing a significant advantage over traditional varactor diodes.

Considering that the cooling power of a typical cryogenic refrigerator's 4~K stage is approximately 200~mW, and its MC stage provides about $400 \, \mu\text{W}$ at 100~mK, it is feasible to place up to 20,000 microwave sources on the 4~K stage and 400 on the MC stage. These numbers are promising for realizing a scalable quantum computer; nevertheless, further improvements are necessary. In principle, lower power consumption of tunnel diodes can be fabricated by making the surface area of the tunnel diode smaller. While this would also lead to a reduction in output power, it can be compensated by adjusting the ratio of the number of turns in the spiral inductor to the number of turns in the pickup coil. However, the impact of back action on the oscillator must be carefully considered to ensure reliable operation. Furthermore, considering the need to apply a magnetic field for electron spin qubits, improvements such as using magnetic field-resilient superconducting materials like NbTiN instead of Nb and designing spiral inductors with these materials could be explored.

\section{Conclusion}\label{sec:concl}
In conclusion, we developed a 140~MHz tunnel diode oscillator with low power consumption of 1~$\mu$W and characterized its performance on the 10 mK stage. The achieved output power and frequency fall within the range typically used for qubit readout of electrons in semiconductors used as qubits. The amplitude stability is better than that of the commercial microwave sources measured in this study, which is encouraging for its application in qubit readout.

\backmatter

%\bmhead{Supplementary information}

%If your article has accompanying supplementary file/s please state so here. 

%Authors reporting data from electrophoretic gels and blots should supply the full unprocessed scans for key as part of their Supplementary information. This may be requested by the editorial team/s if it is missing.

%Please refer to Journal-level guidance for any specific requirements.

\bmhead{Acknowledgements}

This work was supported by the RIKEN-Hakubi program, RIKEN Center for Quantum Computing, JST-FOREST, and Yazaki Foundation. We thank Denis Konstantinov and Daimo Yoshikawa for useful discussions.

\bmhead{Data  Availability}

The datasets generated  during and/or  analysed during  the  current  study  are  available from the corresponding author on reasonable request.

\begin{appendices}

\renewcommand{\thefigure}{\arabic{figure}}
\setcounter{figure}{5}
\renewcommand{\theequation}{\arabic{equation}}
\setcounter{equation}{2}

\section{Individual differences in BD-6 tunnel diodes}\label{secA:Indiv_diff}

\begin{figure}[H]
\centering
\includegraphics[width=\linewidth]{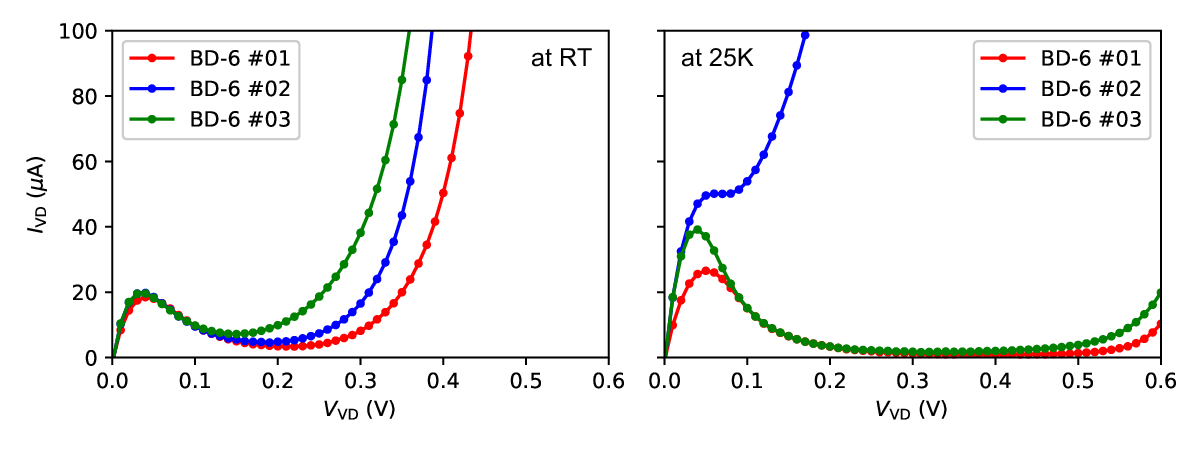}
\caption{I-V curves of three different BD-6 tunnel diodes measured at RT and 25 K. The dots represent measurement points, and the lines connect these points to guide the eye. This measurement was performed using bare tunnel diodes, not those connected to an LC resonator for oscillation. As a result, no oscillation occurs, and thus no hysteresis is observed, unlike in Fig.~\ref{fig:circuit}(d)}
\label{fig:Indiv_diff}
\end{figure}

Fig.~\ref{fig:Indiv_diff} compares the I-V curves of different BD-6 tunnel diodes measured at 25~K and RT. As stated in the BD-6 data sheet, the peak current at RT is 20~$\mu$A, which is consistent across all BD-6 diodes. However, at 25~K, significant variations among individual devices become apparent. Notably, BD-6 \#02 does not exhibit negative resistance at low temperatures, meaning that it cannot be used for the cryogenic microwave source. The BD-6 diode used in this study exhibits a low-temperature I-V characteristic similar to that of BD-6 \#03 but is a different individual device. Typically, a temperature of 25~K is low enough to evaluate and select a suitable BD-6 tunnel diode.

\section{Temperature dependence}\label{secA1}

We conducted detailed measurements of the temperature dependence of the oscillation frequency using a similar circuit. This circuit utilized a BD-6 diode with a parasitic capacitance of 1.7~pF and a 50-turn Nb spiral inductor with an inductance of 1.28~$\mu$H, and did not include a varactor diode. The resulting oscillation frequency was 107~MHz. The circuit described in this work exhibited a comparable temperature dependence based on preliminary measurements.

This circuit is thermally anchored to the 10~mK stage, or the Mixing Chamber (MC) stage, and the MC stage temperature (MCT) was controlled while measuring a thermometer that is also thermally anchored to the MC stage. The sensitivity was found to be 0.3~kHz/mK in the MCT range of 60~mK to 120~mK, with no detectable dependence below 60~mK, aside from a slight hysteresis. This result is significantly better than the temperature dependence reported in previous studies above 1~K~\cite{Van_Degrift1975-sl}.  

The observed temperature dependence between 60~mK and 120~mK is unlikely to be due to changes in the tunnel diode's I-V curve, as no significant variation in its characteristics was observed in this range. Therefore, it is improbable that frequency shifts result from a change in the operating point. In prior studies measuring above 1~K~\cite{Van_Degrift1975-sl}, temperature dependence was attributed to the surface impedance, thermal expansion, and magnetic susceptibility induced by iron impurities in the copper used in the inductor. In contrast, this work uses a superconducting material, Nb, with a critical temperature $T_c$ of 6 K, in the inductor. Given that the measured temperature range is significantly below $T_c$, it is unlikely that the temperature dependence is primarily due to changes in the properties of Nb, though some contributions cannot be ruled out. Alternatively, the temperature dependence might arise from changes in the physical properties of the metals used in the PCB substrate.

Moreover, we note that we did not directly measure the temperature of the circuit itself, and its actual temperature may differ from MCT. The absence of detectable temperature dependence below 60~mK may be due to the fact that the diode’s temperature is likely saturated by the liberated heat and does not cool further.  Please also note that the TDO circuit described in the main text is similarly thermally anchored to the MC stage, and its temperature was measured using a thermometer separately thermally anchored to the MC stage. However, while the qubit must be cooled to 10~mK, the TDO circuit does not necessarily need to be at 10~mK. For the purposes discussed in this work, this is not a concern.

\section{Varactor diode IV curve}\label{secA:varactor}

We measured the IV curve of the varactor diode at 11~mK. The range in which the leak current is suppressed matches the range in which the output power remains constant (Fig.~\ref{fig:freq_tunability}).

\begin{figure}[H]
\centering
\includegraphics[width=0.5\linewidth]{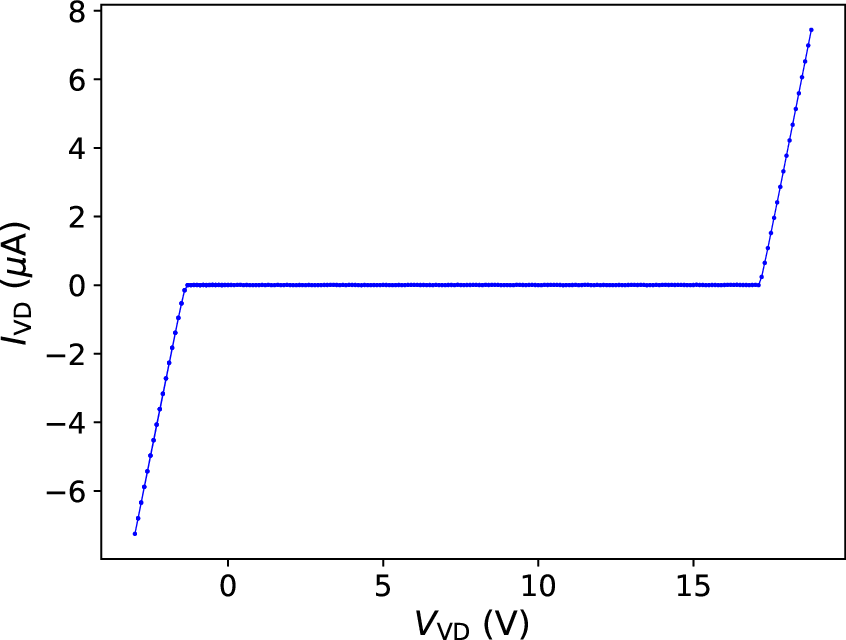}
\caption{IV curve of the varactor diode  (MA46H201, MACOM) measured at 11 mK. For $-1.3~V<V_\mathrm{VD}<17~V$, the leakage current is significantly small.}
\label{fig:varactor}
\end{figure}

\section{DC voltage source stability}\label{secA:DC_source}

\begin{figure}[H]
\centering
\includegraphics[width=0.9\linewidth]{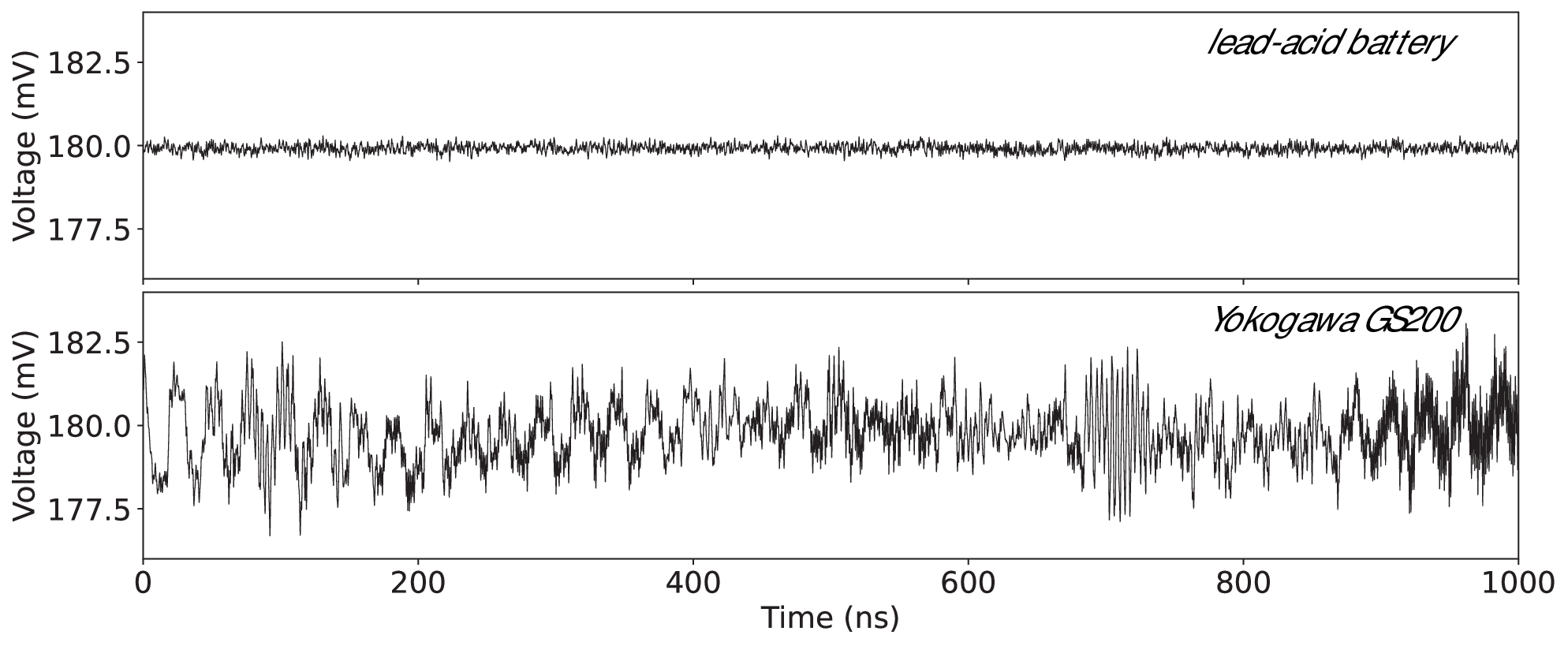}
\caption{Comparison of voltage stability between the Yokogawa GS200 and a lead-acid battery}
\label{fig:DC_comparasion}
\end{figure}

We discovered that the DC voltage source (Yokogawa GS200) was significantly affected by strong electromagnetic interference from a nearby radio station broadcasting at 810~kHz. The impact of the 810~kHz signal was directly observed in the output voltage of the Yokogawa GS200 (Fig.~\ref{fig:DC_comparasion}) and also in the frequency stability of the TDO.

Despite attempts to mitigate the interference using ferrite coils, shielding the lines used for delivering the DC voltage, and shielding the laboratory where the experiment was conducted, the 810~kHz radio signal in the lab was too strong to be removed, likely due to the radio station being only 200~m away. As a result, we decided to switch to using a lead-acid battery, which is not affected by the 810~kHz radio signal. Consequently, as shown in Fig.~\ref{fig:phase_noise}, the phase noise of the TDO was significantly improved.

\section{Time-domain data}\label{secA:time-domain}

\begin{figure}[H]
\centering
\includegraphics[width=\linewidth]{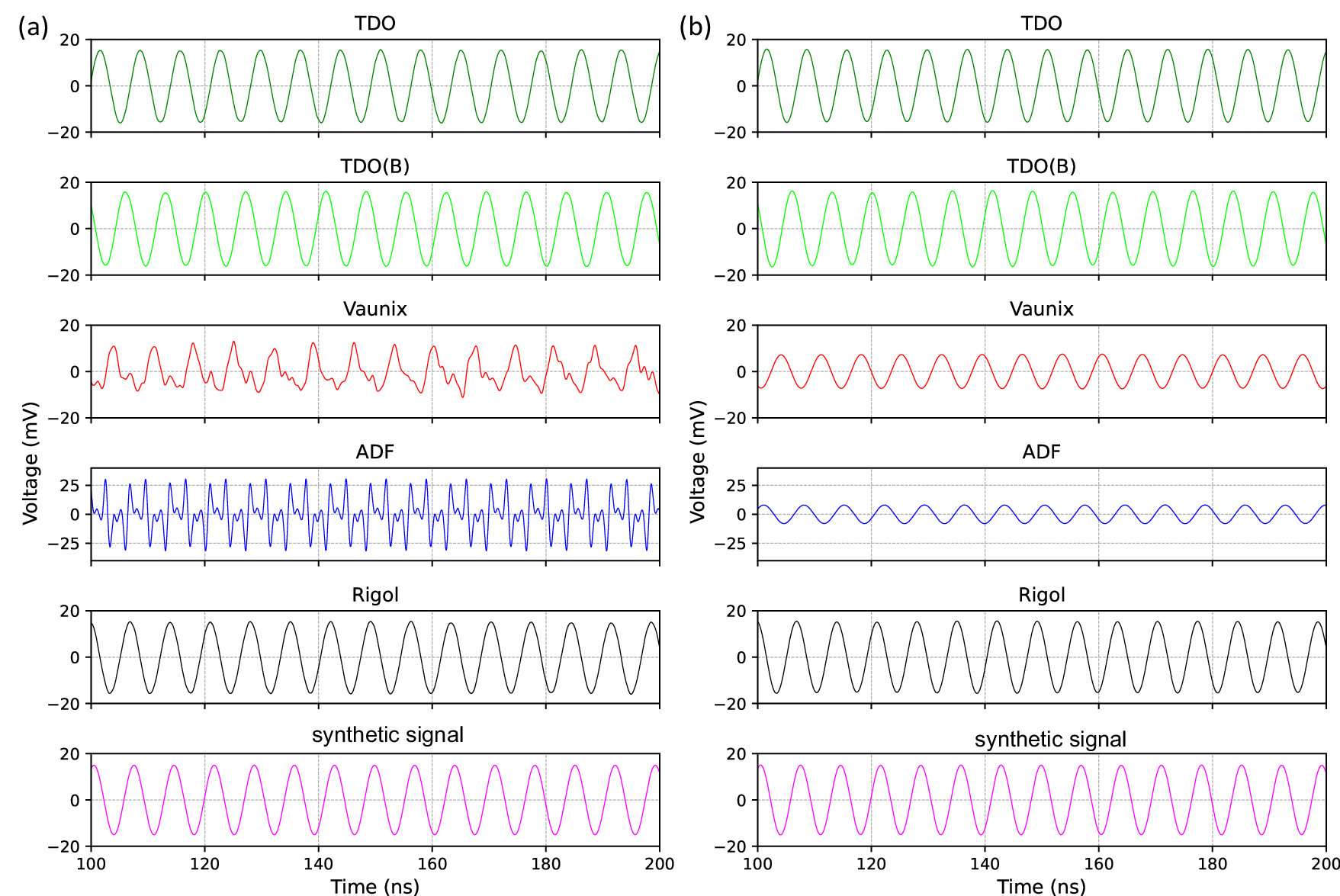}
\caption{Time-domain data measured via AC port II using a 20~GS/s sampling rate oscilloscope (LeCroy WaveRunner 9054). The data include signals from the TDO powered by Yokogawa GS200 DC source (green) and a lead-acid battery (light green), as well as four different commercial microwave sources: Rigol DSA815 (black), Analog Devices ADF4351 (blue), and Vaunix Lab-Brick LMS-451D (red). The respective carrier frequencies are $f_0=$141.800~MHz, 141.841~MHz, 141.802~MHz, 141.806~MHz, and 141.803~MHz. For the TDO, $V_\mathrm{TD}=0.18$~V, and $V_\mathrm{VD}=-0.5$~V. For a consistency check, a synthetic sine signal at 141.8~MHz with 15~mV amplitude is also plotted. The raw signal is shown in (a), while (b) presents the signal after applying a digital bandpass filter with a 50~MHz bandwidth centered on the carrier frequency.}
\label{fig:time-domain}
\end{figure}

The time-domain data were sampled at 20~GS/s using an 8-bit oscilloscope over a duration of 1.6~ms. 100~ns segments are shown in Fig.~\ref{fig:time-domain}. The signals from the ADF and Vaunix generators contain many harmonics, but their influence is eliminated after applying a digital filter (Fig.~\ref{fig:time-domain}). Such harmonics could become an issue when generating signals for multiplexing using these generators. On the other hand, the signal from the TDO shows no observable harmonics, giving it an advantage in this regard.

\section{Power spectra}\label{secA:power_spect}

\begin{figure}[H]
\centering
\includegraphics[width=\linewidth]{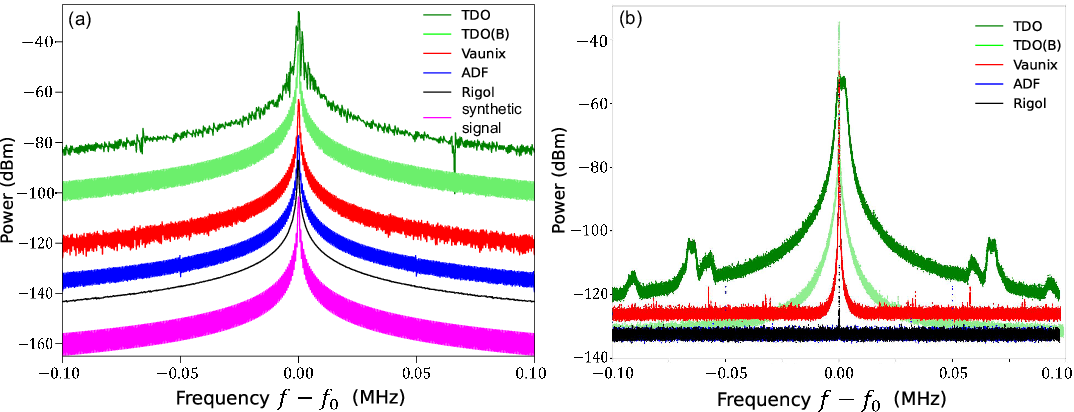}
\caption{(a) Power spectra calculated from the time-domain signal measured at 20~GS/s over a duration of 1.6~ms (see Fig.~\ref{fig:time-domain}). The carrier frequency is the same as in Fig.~\ref{fig:time-domain}. For clarity, 15 dB offsets have been introduced, as the original data are at similar dBm levels. (b) Power spectra measured using a spectrum analyzer (R\&S FSV3030) with a resolution bandwidth of 5~Hz. The carrier frequencies are $f_0 = $ 141.80~MHz and 141.86~MHz for the TDO powered by Yokogawa GS200 DC source (green) and a lead-acid battery (light green), respectively, and $f_0 = $ 141.80~MHz, 141.81~MHz, and 141.80~MHz for the Rigol DSA815 (black), Analog Devices ADF4351 (blue), and Vaunix Lab-Brick LMS-451D (red), respectively. In both (a) and (b), the x-axes represent the frequency offset from the center carrier frequency $f_0$.}
\label{fig:power_spectra}
\end{figure}

Fig.~\ref{fig:power_spectra}(a) shows the power spectra obtained from the time-domain signal in Fig.~\ref{fig:time-domain} with an integration time of 1.6~ms, while Fig.~\ref{fig:power_spectra}(b) shows the power spectra measured using a spectrum analyzer with an RBW of 5~Hz. In Fig.~\ref{fig:power_spectra}(a), little difference is observed between the TDO, commercial microwave sources, and even the synthetic signal. This is because the high-frequency phase noise (see Fig.~\ref{fig:phase_noise}) is low and relatively similar for all cases. The differences observed in Fig.~\ref{fig:power_spectra}(b) must originate from variations in low-frequency phase noise.

To obtain the power spectra shown in Fig.~\ref{fig:power_spectra}(a), the discrete Fourier transform with zero-padding was computed using the following formula:

\begin{equation}
    X_k = \sum_{n=0}^{N-1} x_n \exp\left( -j 2 \pi \frac{nk}{N} \right),
\end{equation}
where the time-domain data $x_n$ is shown in Fig.~\ref{fig:time-domain}(b) and 
$X_k$ represents the Fourier coefficients. The root-mean-square amplitude in volts was calculated as:

\begin{equation}
    \bar{X_k} = \frac{1}{\sqrt{2}} \frac{|X_k|}{N/2},
\end{equation}
where $|X_k|$ is the magnitude of the Fourier coefficient and $N$ is the total number of samples. The power spectrum in dBm, assuming a 50~$\Omega$ system, was computed using:

\begin{equation}
    P = 10 \log_{10} \left( \frac{\bar{X_k}^2}{0.001 \cdot 50} \right),
\end{equation}
where $\bar{X_k}^2/50$ represents the power in watts.

\end{appendices}

\bibliography{library}% common bib file

\end{document}